\documentclass[usegraphicx]{mn2e}
\usepackage{graphicx,verbatim,amsmath,amssymb,amsfonts}
\usepackage{subfigure}
\usepackage{float}
\usepackage{color}

\begin{document}

\title{Radio recombination lines from obscured quasars with the SKA}
\author[S. Manti, S. Gallerani, A. Ferrara, C. Feruglio, L. Graziani, G. Bernardi]{
S. Manti$^1$, S. Gallerani$^1$, A. Ferrara$^1$, C. Feruglio$^1$, L. Graziani$^2$, G. Bernardi$^{3,4,5}$ %\thanks{E-mail: serena.manti@sns.it}
\\
$^1$Scuola Normale Superiore, Piazza dei Cavalieri 7, 56126, Pisa, Italy\\
$^2$INAF -- Osservatorio astronomico di Roma, via Frascati 33, 00040 Monte Porzio Catone, Italy\\
$^3$SKA SA, 3rd Floor, The Park, Park Road, Pinelands, 7405, South Africa\\
$^4$Department of Physics and Electronics, Rhodes University, Grahamstown, South Africa\\
$^5$Harvard-Smithsonian Center for Astrophysics, Cambridge, MA, USA
}
\date{\today} 
\pagerange{\pageref{firstpage}--\pageref{lastpage}} \pubyear{2015} 
\maketitle

\setlength\parindent{12pt}

\begin{abstract}
We explore the possibility of detecting hydrogen radio recombination lines from $0<z<10$ quasars. We compute the expected H$n\alpha$ flux densities as a function of absolute magnitude and redshift by considering (i) the range of observed AGN spectral indices from UV to X-ray bands, (ii) secondary ionizations from X-ray photons, and (iii) stimulated emission due to nonthermal radiation. All these effects are important to determine the line fluxes. We find that the combination of slopes: $\alpha_{\text{X,hard}}=-1.11$, $\alpha_{\text{X,soft}}=-0.7$, $\alpha_{\text{EUV}}=-1.3$, $\alpha_{\text{UV}}=-1.7$, maximizes the expected flux, $f_{\text{H}n\alpha}\approx 10$ $\mu$Jy for $z\sim 7$ quasars with $M_{\text{AB}}=-27$ in the $n\sim 50$ lines; allowed SED variations produce variations by a factor of 3 around this value. Secondaries boost the line intensity by a factor of 2 to 4%(when compared with the flux density of RRLs arising from galaxies, the expected quasar fluxes with the inclusion of secondary ionizations turn out to be 4 to 13 times greater)
, while stimulated emission in high-$z$ quasars with $M_{\text{AB}}\approx -26$ provides an extra boost to RRL flux observed at $\nu\sim 1$ GHz if recombinations arise in H{\small II} regions with $T_e\approx 10^{3-5}$ K, $n_e\approx 10^{3-5}$ cm$^{-3}$. We compute the sensitivity required for a 5$\sigma$ detection of H$n\alpha$ lines using the SKA, finding that the SKA-MID could detect sources with $M_{\text{AB}}\lesssim -27$ ($M_{\text{AB}}\lesssim -26$) at $z \lesssim 8$ ($z\lesssim 3$) in less than 100 hrs of observing time. These observations could open new paths to searches for obscured SMBH progenitors, complementing X-ray, optical/IR and sub-mm surveys. 
\end{abstract}


\section{Introduction}
\hspace{12pt}In the last decade tens of quasars at $z \sim 6$ have been discovered through various different surveys, starting from the seminal work by Fan et al. (2001a, 2001b, 2001c, 2003, 2006a, 2006b), and then continued by Willott and collaborators (2005a, 2005b, 2009, 2010a, 2010b). Nowadays, new surveys as VIKING and PanSTARRS are discovering even more distant quasars (Venemans et al. 2013; Ba$\tilde{\text{n}}$ados et al. 2014). Follow-up observations of emission lines such as the MgII line have shown that these high-redshift quasars (with luminosities well in excess of $10^{47}$ erg $\text{s}^{-1}$) are powered by super-massive black holes (SMBHs) rapidly grown up to a mass $M_{\bullet} \sim (0.02-1.1)\times10^{10} M_{\odot}$ (Barth et al. 2003; Priddey et al. 2003; Willott et al. 2003; Willott et al. 2005a; Jiang et al. 2007; Wang et al. 2010; Mortlock 2011; Wu et al. 2015) in less than a billion years, depending on a still uncertain initial black hole seed (Haiman 2004; Shapiro 2005; Volonteri $\&$ Rees 2006; Tanaka $\&$ Haiman 2009; Treister 2013; Lupi et al. 2014; Tanaka 2014). 

This evidence raises several problems about the formation process and growth of these compact objects (Rees 1978; Volonteri 2010; Latif et al. 2013). Several mechanisms are thought to produce SMBH ancestors, such as the collapse of the first generation of stars (PopIII stars) (Tegmark et al. 1997; Madau $\&$ Rees 2001; Palla et al. 2002; Shapiro 2005), gas dynamical instabilities (Haehnelt $\&$ Rees 1993; Loeb $\&$ Rasio 1994; Eisenstein $\&$ Loeb 1995; Bromm $\&$ Loeb 2003; Koushiappas et al. 2004; Begelman et al. 2006; Lodato $\&$ Natarajan 2006), stellar dynamical processes (Begelman $\&$ Rees 1978; Ebisuzaki et al. 2001; Miller $\&$ Hamilton 2002; Portegies Zwart $\&$ McMillan 2002; Portegies Zwart et al. 2004; G$\ddot{\text{u}}$rkan et al. 2004, 2006; Freitag et al. 2006a, 2006b) and even direct collapse from the gas phase (Haehnelt $\&$ Rees 1993; Begelman et al. 2006; Petri et al. 2012; Yue et al. 2013; Ferrara et al. 2014; Yue et al. 2014; Pallottini et al. 2015).

However, the processes responsible for the growth of initial black holes are still uncertain, also because the ancestors of SMBHs, namely sources with black hole masses of $10^{6-8} M_{\odot}$ at $z>7$, have never been detected so far in spite of the enormous progresses produced by deep X-ray and IR surveys. One possibility hinted by several studies (Yue et al. 2013), including the most recent one by Comastri et al. (2015), envisages that, before the quasar mode is on, accreting black holes are enshrouded by a thick cocoon of gas (and dust) heavily absorbing their optical/X-ray radiation. These sources would then escape detection by the standard X-ray surveys. 

The transparency of gas and dust to radio, millimetre and submillimetre photons makes these kind of observations a promising tool for detecting obscured star formation, quasars in general, and, in particular, super massive black holes progenitors at very early epochs (e.g. Spaans \& Meijerink 2008; Gallerani et al. 2012; Gallerani et al. 2014). Here, we investigate the possibility to detect high-redshift ($z > 7$) quasars that might be completely obscured at optical to X-ray wavelengths by means of hydrogen Radio Recombination Lines (RRLs; for a comprehensive review, see Gordon \& Sorochenko 2002). In fact, one advantage of RRLs is that they are nearly always in the optically thin limit and therefore see all the gas. The other advantage of using hydrogen RRLs concerns with the fact that this method could work equally well for gas of primordial composition, as it relies only on the presence of hydrogen rather than lines from heavy elements.

Hydrogen RRLs represent a special class of spectral lines arising in HII regions from transitions between highly excited hydrogen levels (quantum numbers $n>27$) and appearing in the radio regime (rest frame frequencies $\nu_e<300$ GHz); H$n\alpha$ lines are the strongest hydrogen RRLs, i.e. those due to $n+1 \rightarrow n$ transitions. Also note that RRLs have the significant advantage that the principal quantum number of a line detectable at a fixed observed frequency decreases with increasing redshift. Given that the RRL flux density is proportional to $(1+z)(d_L(z))^{-2}n^{-2.72}$ (Rule et al. 2013, hereafter R13), this effect partially compensates the increase of the luminosity distance $d_L$ with redshift.

The search for RRLs from obscured quasars will become feasible in the near future with the SKA (Square Kilometre Array) telescope, thanks to its revolutionary capabilities in terms of frequency range, angular resolution and sensitivity (for SKA science and technical specifications, see: Morganti et al. 2014; SKA-­‐SCI-­‐LVL-­‐001, 2015\footnote{http://astronomers.skatelescope.org/wp-content/uploads/2015/08/SKA-SCI-LVL-001W.pdf
%https://www.skatelescope.org/wp-content/uploads/2014/03/SKA-OFF.SE$_{}$.ARC-SKO-SRS-001$_{}$3$_{}$Level$_{}$1$_{}$Requirements-3-signed.pdf.
}). In particular, the SKA-MID (0.35--14 GHz) frequency coverage will allow us to search for H$n\alpha$ ($35 < n < 210$) lines from sources at $1 < z < 10$. For all these reasons, in principle, the detectability of high-$z$ obscured quasars through RRLs could allow us to directly monitor the birth and growth of the earliest quasars during the first cosmic billion years.

The paper is organized as follows. In Sec. 2, we derive the RRL flux density expected from quasars, including secondary ionizations from X-ray photons. In Sec. 3, we test our model against available RRL observations. In Sec. 4, we analyse the contribution of the stimulated emission due to a nonthermal background source to the quasar RRL flux. In Sec. 5, we discuss the prospects for detecting obscured quasars through the SKA. In Sec. 6 we discuss the main conclusions of our work.

\begin{figure}%[H]
\centering
\hspace{-0.8cm}
\includegraphics[width=0.5\textwidth]{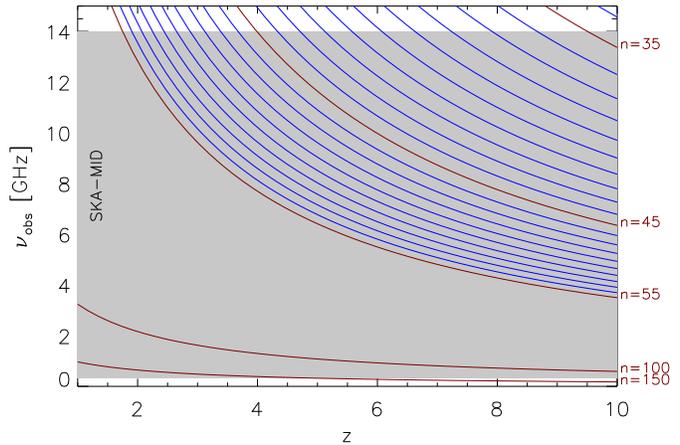}
\caption{Hydrogen RRL observed frequencies (in GHz) as a function of redshift. Blue curves correspond to quantum numbers from $n=34$ to $n=54$, and red lines correspond to $n=35$, $n=45$, $n=55$, $n=100$ and $n=150$. The lines with $n=56$ to $149$ are not shown for the sake of clarity. The grey region shows the frequency range covered by SKA-MID (0.35--14 GHz).}
\label{nuobs_z}
\end{figure}

\section{RRL\lowercase{s} from quasars}
\hspace{12pt}The observed frequencies of H$n\alpha$ lines are given by the following equation:
\begin{equation}
\nu_{\text{obs}}(z_e,n)=\frac{c\hspace{0.05cm}\text{R}_{\text{H}}}{1+z_e}\left[\frac{1}{n^2}-\frac{1}{(n+1)^2}\right],
\label{nu_obs}
\end{equation}
where $z_e$ is the redshift at which the line is emitted, $c$ is the speed of light, $\text{R}_{\text{H}}=1.0968 \times 10^5$ $\text{cm}^{-1}$ is the Rydberg number for hydrogen, and $n$ is the principal quantum number. In Fig. \ref{nuobs_z}, we show the observed line frequencies as a function of redshift for different values of $n$; the frequency range covered by SKA-MID is shown through a grey shaded region.
 
The flux density at the center of H$n\alpha$ lines ($f_{\text{H}n\alpha}$) can be written as (R13) %(Rule et al. 2013, hereafter R13)
\begin{equation}
f_{\text{H}n\alpha}\approx 3.25\hspace{0.05cm}n^{-2.72}\dot{N}_{\gamma}\hspace{0.05cm}(1-f_{\text{esc}}^{912})\hspace{0.05cm}\frac{hc}{\delta v}\hspace{0.05cm}\frac{(1+z_e)}{4\pi d_L^2},
\label{main_formula}
\end{equation}
where $d_L$ is the luminosity distance of an emitting source at redshift $z_e$, $\delta v$ is the width of the line in velocity units, $h$ is the Planck constant, and $f_{\text{esc}}^{912}$ is the escape fraction of ionizing photons. Since we want to test the possibility of detecting optically obscured quasars through RRLs, we assume that all the ionizing photons remain trapped into a dense, fully neutral, surrounding medium, namely $f_{\text{esc}}^{912}=0$. We note that Eq. (\ref{main_formula}) is valid for $\alpha$-transitions within the interval $n=3-29$ (Hummer $\&$ Storey 1987), with an accuracy of a few percent for any value of the electron density $n_e$ (R13). We assume that this equation can be used even for higher values of $n$.
%{\bf Eq. (\ref{main_formula}) takes an approximation based on the results for $\alpha$-transitions within the interval $n=3-29$, with an accuracy of a few percent (Hummer $\&$ Storey 1987). Since our aim is to only roughly estimate the line flux densities from quasars, this equation can be used even for higher values of $n$, with an accuracy of...}

Finally, the production rate of ionizing photons $\dot{N}_{\gamma}$ is given by
\begin{equation}\label{ndot1}
\dot N_{\gamma}= \int _{\nu_{\text{LL}}}^{+\infty} \frac{L_\nu}{h\nu} d\nu ,
\end{equation}
where $\nu_{\text{LL}}$ is the Lyman limit frequency ($h\nu_{LL}=13.6$ eV) and $L_{\nu}$ is the luminosity per unit frequency. Substituting the usual relations:
\begin{equation}
L_{\nu}=4\pi d_L^2f_{\nu} ; \hspace{0.5cm}\lambda f_{\lambda}=\nu f_{\nu} ,
\end{equation}
we can rewrite eq. (\ref{ndot1}) as
\begin{equation}
\dot N_{\gamma}=\frac{4\pi}{hc}\int_0^{\lambda_{\text{LL}}}d_L^2(z)\lambda f_{\lambda} d\lambda ,
\label{Ndot}
\end{equation}
where $\lambda_{\text{LL}}=c/\nu_{\text{LL}}=911.8\hspace{0.1cm}\text{\AA}$. This equation clearly shows that $\dot N_{\gamma}$ (and therefore $f_{\text{H}n\alpha}$) depends on the Spectral Energy Distribution (SED) of the ionizing source.

\begin{figure}%[H]
\centering
\hspace{-0.9cm}
\includegraphics[width=0.58\textwidth,height=0.285\textheight,keepaspectratio]{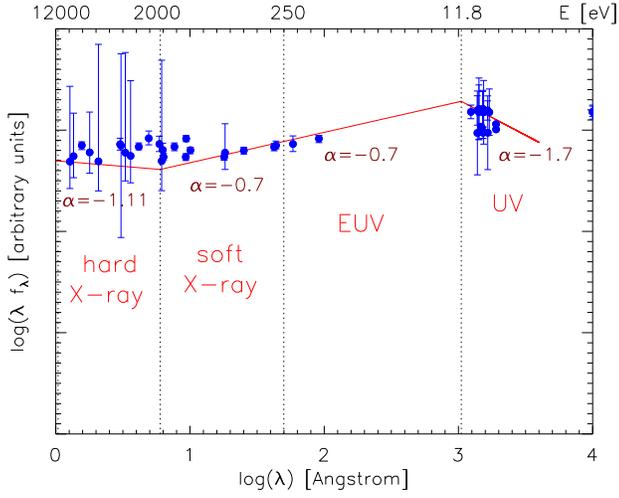}
\caption{Schematic plot of the quasar SED, referring to our ``fiducial'' case with $\alpha_{\text{X,hard}}=-1.11, \alpha_{\text{X,soft}}=-0.7, \alpha_{\text{EUV}}=-0.7, \alpha_{\text{UV}}=-1.7$. Blue points represents measurements of the quasar NGC 1068. The upper $x$-axis shows the extreme energy values of each band.}
\label{SED_figure}
\end{figure}

\subsection{Quasar SED}\label{SED}
\hspace{12pt} 
We parametrize the emitted quasar flux per unit wavelength as $f_{\lambda} \propto \lambda^{\alpha}$, where the spectral index $\alpha$ depends on the wavelength range. We adopt the following observationally constrained $\alpha$ values in different energy bands: 
\begin{itemize}
\item $\alpha_{\text{X,hard}}=-1.11 \pm 0.11$ for the hard X-ray band (2500--12000 eV, Piconcelli et al. 2005); 
\item $-0.7 < \alpha_{\text{X,soft}} < 0.3$ for the soft X-ray band (250--2000 eV, Fiore et al. 1994); 
\item $\alpha_{\text{EUV}}=-0.7^{+0.8}_{-0.6}$ for the EUV band (12--200 eV, Wyithe \& Bolton 2011); 
\item $\alpha_{\text{UV}}=-1.7^{+1.5}_{-0.9}$ for the UV band (5.5--11.8 eV, Reichard et al. 2003). 
\end{itemize}
Fig. \ref{SED_figure} presents the quasar SED obtained in our ``fiducial'' case, given by the following combinations of slopes: $\alpha_{\text{X,hard}}=-1.11, \alpha_{\text{X,soft}}=-0.7, \alpha_{\text{EUV}}=-0.7, \alpha_{\text{UV}}=-1.7$. The blue points represent the photometric data of the quasar NGC 1068 (from NASA/IPAC Extragalactic Database (NED)). This example show how our adopted spectral indices, once compared with the observations, well reproduce real measurements of AGN spectra. Afterwards, we will also consider the case in which $\alpha_{\text{X,hard}}=-1.11, \alpha_{\text{X,soft}}=-0.7, \alpha_{\text{EUV}}=-1.3, \alpha_{\text{UV}}=-1.7$, which we call ``extreme'' model.

\begin{figure}%[H]%[ht!]
\vspace{-1.1cm}
\centering 
 \begin{tabular}{@{}cc@{}}
        \hspace{-0.5cm}\subfigure{\includegraphics[width=0.5\textwidth]{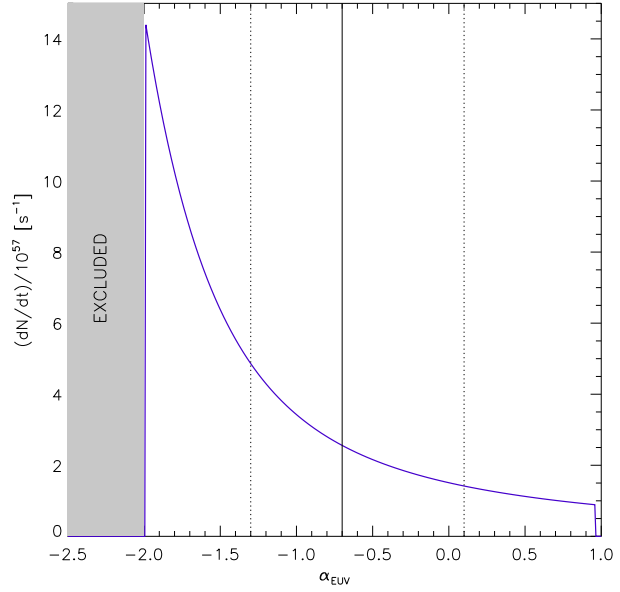}}
        \vspace{-1cm}\\
        \hspace{-0.58cm}\subfigure{\includegraphics[width=0.5\textwidth]{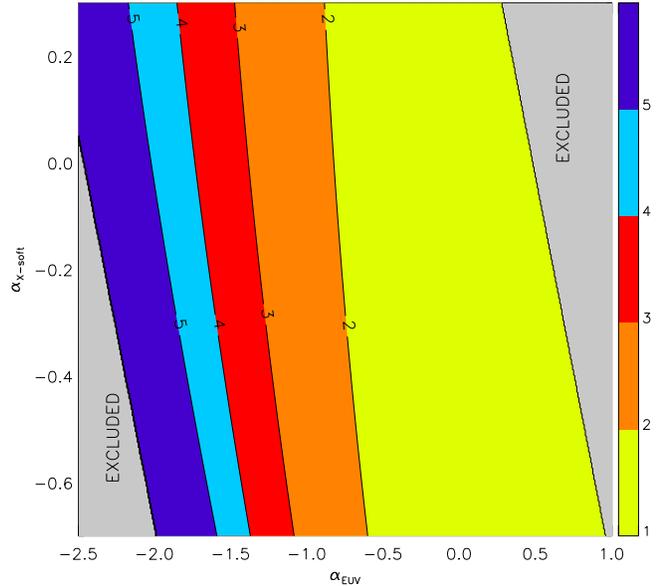}}
\end{tabular} 
\vspace{-0.5cm} 
\caption{Upper: ionization rate in units of $10^{57}$ s$^{-1}$ versus $\alpha_{\text{EUV}}$, with fixed values $\alpha_{\text{X,hard}}=-1.11$, $\alpha_{\text{X,soft}}=-0.7$, $\alpha_{\text{UV}}=-1.7$, for a quasar magnitude $M_{\text{AB}}^{1500}=-27$. The vertical lines indicate the best value (solid) and the 2$\sigma$ C.L. values (dotted) of $\alpha_{\text{EUV}}$. Lower: contour plot of the boost due to secondary ionizations as a function of $\alpha_{\text{X,soft}}$ and $\alpha_{\text{EUV}}$, with $\alpha_{\text{X,hard}}=-1.11$ and $\alpha_{\text{UV}}=-1.7$. Contours correspond to $\dot{N}_{\text{Q}}^{\text{sec}}/\dot{N}_{\text{Q}}=1, 2, 3, 4, 5$. In both panels, the `excluded' regions (in grey) are the ones in which the constraint $-2<\alpha_{ox}<-0.5$ does not hold.} 
\label{contours}
\end{figure}

We compute the production rate of ionizing photons in quasars ($\dot{N}_{Q}$) by substituting the SED in eq. (\ref{Ndot}):
\begin{align}
\dot N_{Q} &=\frac{4\pi d_L^2(z)}{hc}f_{1500} \bigg[f_{\text{X,hard}}\int_{\lambda_{\text{min}}}^{\lambda_{2\text{ keV}}}\lambda\left(\frac{\lambda}{\lambda_{2\text{ keV}}}\right)^{\alpha_{\text{X,hard}}} d\lambda \nonumber\\
&+ f_{\text{X,soft}}\int_{\lambda_{2\text{ keV}}}^{\lambda_{0.25\text{ keV}}}\lambda\left(\frac{\lambda}{\lambda_{0.25\text{ keV}}}\right)^{\alpha_{\text{X,soft}}} d\lambda \nonumber\\
&+ f_{\text{EUV}}\int_{\lambda_{0.25\text{keV}}}^{\lambda_{\text{LL}}}\lambda\left(\frac{\lambda}{1050\hspace{0.1cm}\text{\AA}}\right)^{\alpha_{\text{EUV}}} d\lambda \bigg].
\label{Ndot-final}
\end{align}
Here $\lambda_{\text{min}}=hc/E_{\text{max}}$ ($E_{\text{max}}=10$ keV), $f_{\text{X,hard}}$, $f_{\text{X,soft}}$ and $f_{\text{EUV}}$ are normalization factors required to match the flux densities between different bands of the spectrum and $f_{1500}$ is the flux density at 1500 $\text{\AA}$, related to the AB magnitude ($M_{\text{AB}}^{1500}$) through the following relation (R13)
\begin{equation}
f_{1500}=\frac{c}{\lambda^2}10^{-(M_{\text{AB}}^{1500}+5\log_{10}d_L(z)+43.6)/2.5}.
\label{norm1500}
\end{equation}  
The UV spectral index, $\alpha_{\text{UV}}$, enters in eq. (\ref{Ndot-final}) through the normalization $f_{\text{EUV}}$.

While varying the spectral indices in the different energy bands in the allowed intervals, we compute the spectral hardness between the UV and X-ray bands (Tananbaum et al. 1979) defined as  
\begin{equation}
\alpha_{ox}=0.3838\log[L_{2\hspace{0.05cm}\text{keV}}/L_{2500\hspace{0.05cm}\text{\AA}}],
\end{equation}
where $L_{2\hspace{0.05cm}\text{keV}}$ and $L_{2500\hspace{0.05cm}\text{\AA}}$ are the monochromatic luminosities at 2 keV and 2500 $\text{\AA}$, respectively. Observations allow us to constrain this parameter in the range $-2<\alpha_{ox}<-0.5$ (Wu et al. 2012).

We find that $\dot N_{Q}$ strongly depends on the assumed value of $\alpha_{\text{EUV}}$ (while being completely independent on $\alpha_{\text{X}}$ and $\alpha_{\text{UV}}$), varying between $10^{57}<\dot N_{Q}/\hspace{0.1cm}\text{s}^{-1}<14\times10^{57}$ for $1>\alpha_{\text{EUV}}>-2$, as shown in the top panel of Fig. \ref{contours}, where we fix $M_{\text{AB}}^{1500}=-27$. The grey region represents the $\alpha_{\text{EUV}}$ values excluded by the $\alpha_{\text{OX}}$ constraints. The vertical lines shown in the figure correspond to the $\alpha_{\text{EUV}}$ best value (solid) and to the 2$\sigma$ confidence level (C.L.) values (dotted). 
\subsection{Secondary ionizations}
\label{sec_ion}
\hspace{12pt}
Since quasars are particularly bright in the X-ray wavelength range, the contribution of secondary ionizations from fast electrons should be taken into account in the estimate of $\dot N_{Q}$. In fact, a non negligible fraction $f_i$ of the energy of these electrons ($E_e = h (\nu -\nu_{\text{LL}})$) released by photo-ionization, is expected to cause additional (i.e. secondary) ionizations from collisions on the neutral component. The efficiency of the secondaries is then a function of both the electron energy and the ionized fraction of the gas $x_e$.
According to the model presented by Vald\'{e}s \& Ferrara (2008) (see also Furlanetto \& Stoever 2010; Evoli et al. 2012), in a cosmological pristine gas, the relation between $f_i$ and $x_e$ can be written as\footnote{We assume an initial energy of the primary electron in the range 3-10 keV.} 
\begin{equation}\label{sec_ion}
f_i=0.3846(1.0-x_e^{0.5420})^{1.1952}.
\end{equation}

We compute the production rate of ionizing photons including secondary ionizations ($\dot{N}_{\text{Q}}^{\text{sec}}$) by replacing each factor $\lambda f_{\lambda}$ in eq. (\ref{Ndot}) with $\lambda f_{\lambda}\left(1+\frac{hc}{13.6 \mathrm{eV}\lambda}f_i\right)$. In particular, since we want to compute the effect of secondary ionizations in the fully neutral medium surrounding the H{\small II} region, we assume $x_e\sim 0$. As the secondaries deriving from photo-ionization have a spectral energy distribution, we account for the dependence $f_i(E_e)$ by considering the relation $f_i=4.94\arctan{E_e}-7.37$ for $x_e\sim 0$, obtained by fitting the corresponding curve in Fig. 2 from Furlanetto \& Stoever (2010).

\begin{figure}%[H]
\centering
\hspace{-0.5cm}
\includegraphics[width=0.49\textwidth]{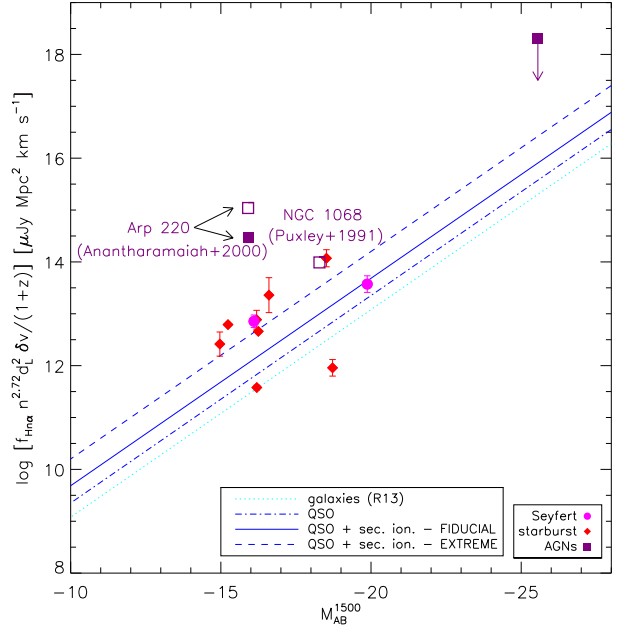}
\caption{Logarithm of $[f_{\text{H}n\alpha} n^{2.72} d_L^{2} \delta v/(1+z)]$ as a function of $M_{\text{AB}}^{1500}$. The observational data of RRL flux densities from extragalactic sources at $0.000811<z<0.158$ with the relative error bars (see Table \ref{table_extragalSources}) % (Allen et al. 1968; Anantharamaiah et al. 1993, 1996, 1997, 1998, 2000; Das et al. 2001; Kepley et al. 2011; Mohan et al. 2001, 2002, 2005; Phookun et al. 1998; Rodr\'{i}guez-Rico et al. 2005; Roy et al. 2005, 2008, 2010; Zhao et al. 1996, 1997) 
are plotted with different colors and symbols: magenta circles represent Seyfert galaxies, red diamonds denote starburst galaxies and purple squares are AGNs (filled and empty are respectively H90$\alpha$/H92$\alpha$ and H53$\alpha$ line observations). The arrow indicates that 3C 273 has an upper limit flux only. For comparison, the lines representing the theoretical fluxes predicted for galaxies from R13 (cyan dotted) and for quasars without (dot-dashed blue, for the ``fiducial'' case) and with the inclusion of secondary ionizations with $x_e=0$ (solid and dashed blue lines, corresponding to our ``fiducial'' and ``extreme'' model respectively) are also shown.}
\label{flux_densities}
\end{figure}

\begin{table*}%[h]
\caption{Local extragalactic sources with the observed line parameters and relative references.}
\centering
%\small
\begin{tabular}{@{}lccccccl}
\hline\hline
source & $z$ & $m_{\text{AB}}^{1500}$ & $n$ & FWHM (km s$^{-1}$) & $F_{\text{peak}}$ (mJy) & source type & references\\
\hline
ARP 220    &    0.0018   &  18.57  & 92 &   $363\pm 45$  &          $0.6\pm 0.1$      &    2 merging AGNs               & (a) \\
ARP 220    &    0.0018   &  18.57  & 53 &   $230\pm 20$  &          $16\pm 2$         &                                 & (b) \\
NGC 3628   &    0.002812 &  14.23  & 92 &   $170\pm 70$  &          $1.4\pm 0.09$     &    starburst                    & (c),(d) \\
IC 694     &    0.013200 &  15.28  & 92 &   $350\pm 110$ &          $0.47\pm 0.10$    &    starburst                    & (c),(d) \\
NGC 1365   &    0.005457 &  11.99  & 92 &   $310\pm 110$ &          $0.99\pm 0.10$    &    Seyfert                      & (c) \\
NGC 5253   &    0.001358 &  12.64  & 92 &   $95\pm 10$   &          $0.53$            &    starburst                    & (e) \\
NGC 660    &    0.002835 &  15.20  & 92 &   $377$        &          $0.5$             &    starburst                    & (f) \\
NGC 3256   &    0.009354 &  16.44  & 92 &   $114\pm 60$  &          $0.56\pm 0.32$    &    starburst                    & (g) \\
NGC 1808   &    0.003319 &  14.68  & 92 &   $339\pm 75$  &          $0.47\pm 0.08$    &    Seyfert                      & (h) \\     
M 83       &    0.001711 &  10.62  & 92 &   $95\pm 30$   &          $0.8\pm 0.15$     &    starburst                    & (i) \\
NGC 2146   &    0.002979 &  15.58  & 92 &   $200\pm 95$  &          $0.36\pm 0.09$    &    starburst                    & (i) \\
NGC 253    &    0.000811 &  11.47  & 92 &   $190\pm 10$  &          $9.0\pm 0.5$      &    starburst                    & (j) \\          
NGC 1068   &    0.003793 &  12.79  & 53 &   $200\pm 100$ &          $37.3\pm 12.9$    &    AGN                          & (k) \\
3C 273     &    0.158    &  13.85  & 90 &   --           &          $< 200$           &    AGN                          & (l) \\
\hline
\hline
\end{tabular}\\
\begin{flushleft}
(a) Anantharamaiah et al. 2000; (b) Rodr\'{i}guez-Rico et al. 2005; (c) Anantharamaiah et al. 1993; (d) Zhao et al. 1997; (e) Mohan et al. 2001; (f) Phookun et al. 1998; (g) Roy et al. 2005; (h) Roy et al. 2008; (i) Zhao et al. 1996; (j) Anantharamaiah $\&$ Goss 1996, 1997; Das et al. 2001; Kepley et al. 2011; Mohan et al.  2002; Mohan et al. 2005; (k) Puxley et al. 1991; (l) Allen et al. 1968
\end{flushleft}
\label{table_extragalSources}
\end{table*}

We quantify the boost of the H$n\alpha$ line flux density due to secondary ionizations through the ratio $\dot{N}_{\text{Q}}^{\text{sec}}/\dot{N}_{\text{Q}}$. In the bottom panel of Fig. \ref{contours}, we plot this quantity as a function of $\alpha_{\text{X,soft}}$ and $\alpha_{\text{EUV}}$. Since $\dot{N}_{\text{Q}}^{\text{sec}}/\dot{N}_{\text{Q}}$ is only slightly dependent on $\alpha_{\text{X,hard}}$ and $\alpha_{\text{UV}}$, we fix these slopes to their best-values ($\alpha_{\text{X,hard}}=-1.11$, $\alpha_{\text{UV}}=-1.7$). As in the top panel, the grey areas represent regions of the parameter space that are excluded by the $\alpha_{\text{OX}}$ constraints. 

We find that $\dot{N}_{\text{Q}}^{\text{sec}}/\dot{N}_{\text{Q}}=2.2$ in the ``fiducial'' case and $\dot{N}_{\text{Q}}^{\text{sec}}/\dot{N}_{\text{Q}}=3.7$ in the ``extreme'' one. Once compared with the production rate of ionizing photons for galaxies, it turns out that $\dot{N}_{\text{Q}}$ ($\dot{N}_{\text{Q}}^{\text{sec}}$) is greater than $\dot{N}_{\text{G}}$ by a factor of 2 (4) to 4 (13), in our ``fiducial'' and ``extreme'' model respectively. For $M_{\text{AB}}^{1500}=-25$, we find $\log \dot{N}_{\text{G}}=56.33$ for galaxies, while for quasars we have $\log \dot{N}_{\text{Q}}=56.60$ and $\log \dot{N}_{\text{Q}}=56.88$ in the ``fiducial'' and ``extreme'' case respectively.

\section{Test with local sources}
\label{tests}
\hspace{12pt}To test the model described above, we compare its predictions with observations of radio recombination lines from a sample of extragalactic sources (starburst galaxies, Seyfert galaxies and AGNs, listed in Table \ref{table_extragalSources}) at $z\sim 0$ (Allen et al. 1968; Anantharamaiah et al. 1993, 1996, 1997, 2000; Das et al. 2001; Kepley et al. 2011; Mohan et al. 2001, 2002, 2005; Phookun et al. 1998; Puxley et al. 1991; Rodr\'{i}guez-Rico et al. 2005; Roy et al. 2005, 2008; Zhao et al. 1996, 1997). In this sample there are two AGNs, NGC 1068 and 3C 273 (the latter has an upper limit flux only). In addition, recent results (Paggi et al. 2013) strongly suggest that Arp 220 also hosts two merging, deeply buried AGNs, whose bolometric luminosity is likely in the range $2\times 10^{42}$ erg $\text{s}^{-1}$ and $5.2\times 10^{43}$ erg $\text{s}^{-1}$ for the eastern and western nuclei, respectively. Finally note that NGC 1068 and Arp 220 are radio-quiet quasars (RQQs) while 3C 273 is a radio-loud source (RLQ).

In this context, the distinction between RQQs and RLQs is important due to the fact that in radio-loud quasars, the presence of a strong radio continuum emission makes possible the detection of RRLs out to greater distances with respect to RQQs, thanks to the stimulation of their line flux densities (see Section 4 for the details).
%, in radio-loud quasars, the strong radio continuum radiation may stimulate the RRL emission thus boosting their flux (Shaver 1978; Shaver et al. 1978). Therefore, in RLQs the RRL flux density is expected to be higher than what we found through our calculations. For this reason, RLQs are in principle detectable out to greater distances with respect to RQQs. 

For each source of the previous sample but NGC 1068 and 3C 273 we extract the observed line parameters of the H92$\alpha$ line ($\nu_{\text{rest}}=8309.383$ MHz). In the case of Arp 220 (besides the H92$\alpha$ detection) and NGC 1068 we have observations of the H53$\alpha$ line ($\nu_{\text{rest}}=42951.97$ MHz) and we use the observed values. Furthermore, a search for the H90$\alpha$ line ($\nu_{\text{rest}}=8872.569$ MHz) in the emission spectrum of 3C 273 has led to an upper limit only in the detected signal. More specifically, we collect the data relative to the quantum number ($n=92$ for all the sources except NGC 1068 and 3C 273, $n=53$ for NGC 1068, both for Arp 220, and $n=90$ for 3C 273), the line width in $\text{km}$ $\text{s}^{-1}$ and the peak line flux density in mJy. 
The redshift $z$ (thus the luminosity distance) and the apparent magnitude $m_{\text{AB}}$ at 1500 $\text{\AA}$ of each source are taken instead from the NED website.

\begin{figure*}%[H]%[ht!]
\centering      
 \begin{tabular}{@{}cc@{}}
       \subfigure{\includegraphics[width=0.5\textwidth]{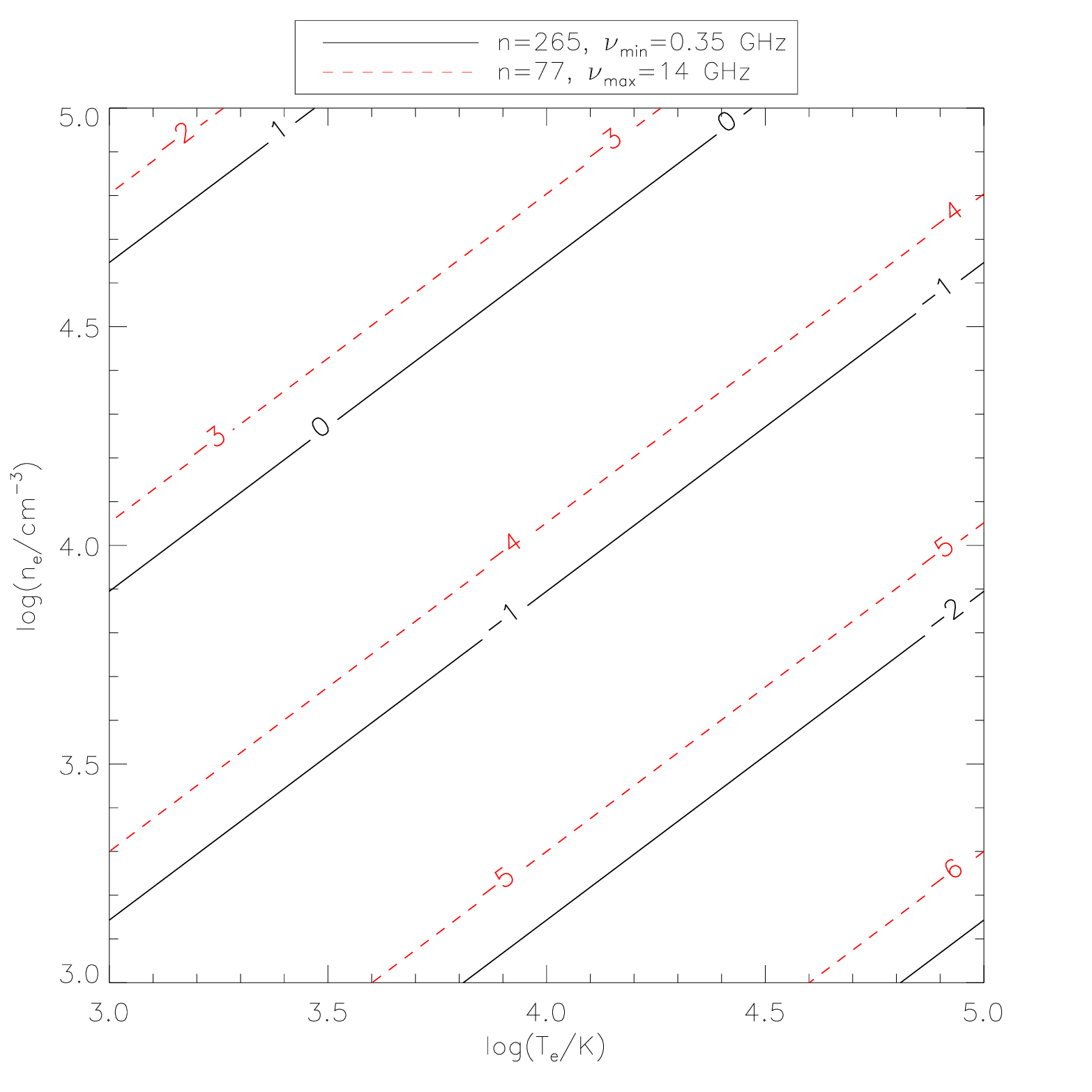}}
       \subfigure{\includegraphics[width=0.5\textwidth]{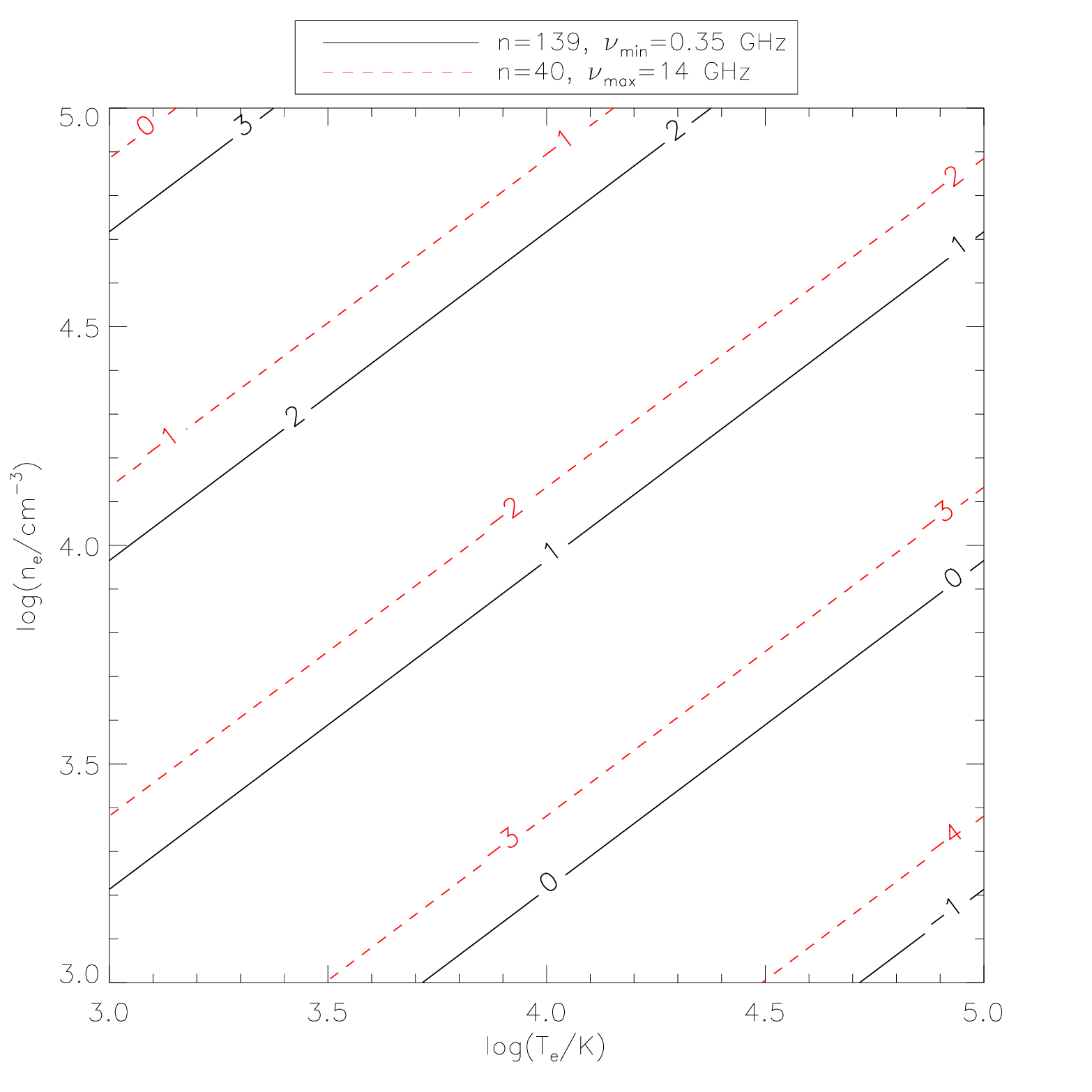}}
%\vspace{-0.5cm}
\\
${\bf z=0}$; ${\bf S_0=10}$ {\bf mJy}.  \hspace{5.5cm} ${\bf z=6}$; ${\bf S_0=0.01}$ {\bf mJy}.
\end{tabular}
\vspace{0.4cm}
\caption{Contour plots of the logarithm of the ratio X between stimulated and spontaneous emission as a function of the temperature $T_e$ and the density $n_e$ of the H{\small II} region, for RRLs of quantum numbers observable at $z=0$ (left panel) and $z=6$ (right panel), with a continuum flux density $S_0=10$ mJy and $S_0=0.01$ mJy respectively. Black solid contours are traced for the minimum SKA-MID frequency, $\nu_{\text{min}}=0.35$ GHz, and correspond to quantum numbers $n=265$ (left) and $n=139$ (right), while red dashed contours are for RRLs detectable in correspondence of the maximum SKA-MID frequency, $\nu_{\text{max}}=14$ GHz, with quantum numbers $n=77$ (left) and $n=40$ (right). The model adopted is our ``fiducial'' case with the inclusion of secondary ionizations, for quasars with magnitude $M_{\text{AB}}^{1500}=-26$.}
\label{contours_z0_z6}
\end{figure*}

Fig. \ref{flux_densities} shows, for this compilation of data, the logarithm of $[f_{\text{H}n\alpha} n^{2.72} d_L^2 \delta v/(1+z)]$ versus $M_{\text{AB}}^{1500}$, with different colors and symbols according to the different class of sources (the downward arrow indicates that the line flux density of 3C 273 is an upper limit). In this figure, we also plot our theoretical predictions: the blue solid line indicates our ``fiducial'' case, the blue dashed line the ``extreme'', and the blue dot-dashed line the ``fiducial'' case without taking into account secondary ionizations. The cyan dotted line represents predictions for the case of low-metallicity ($Z < 10^{-5}$) galaxies without AGN. For these sources, we compute $f_{\text{H}n\alpha}$ according to the R13 formalism:
\begin{align}
f_{\text{H}n\alpha} &\approx 1.2\times 10^{-7}n^{-2.72}10^{-0.4M_{\text{AB}}^{1500}}(1+z) \nonumber\\
&\times \left(\frac{d_L}{10^5 \hspace{0.1cm}\text{Mpc}}\right)^{-2} \left(\frac{\delta v}{100 \hspace{0.1cm}\text{km s$^{-1}$}}\right)^{-1}(1-f_{\text{esc}}^{912}) \hspace{0.1cm}\mu \text{Jy}.
\label{fHnalpha}
\end{align}
In this case, $\dot{N}_{\text{G}}=(2.2\times10^{-11})\times 10^{-0.4M_{\text{AB}}^{1500}}$. Eq. (\ref{fHnalpha}) is obtained from eq. (\ref{main_formula}) by using empirical relations between the star formation rate (SFR) and the 1500 $\text{\AA}$ luminosity (Mu\~{n}oz $\&$ Loeb 2011), and between the SFR and $\dot{N}_{912}$ (Table 4 from Schaerer 2003).

While the R13 model is better suited for comparison with starburst galaxies (red filled diamonds), our predictions should be compared with Seyfert and AGN data (magenta circles and purple squares). We note that the R13 model underpredicts by more than one order of magnitude the observed flux for most of the sources (all but two). Although the $f_{\text{H}n\alpha}$ values predicted by us are $\sim 10$ times higher than the R13 ones, we still can not reproduce the observations of local AGNs. This means that very likely we are neglecting further important contributions to the RRL emission.

\section{Radio stimulated emission}
\hspace{12pt}
Up to now, we have considered only spontaneous emission in radio-quiet quasars. However, radio recombination lines may arise from a combination of spontaneous and stimulated emission (e.g. Bell \& Seaquist 1977). In fact, in radio-loud quasars, the strong radio continuum radiation may stimulate the RRL emission thus boosting their flux (Shaver 1978; Shaver et al. 1978). Therefore, in RLQs the RRL flux density is expected to be higher than what we found through our calculations. For this reason, RLQs are in principle detectable out to greater distances with respect to RQQs. 

In this section, we quantify the relative contribution of the stimulated emission due to a nonthermal background source with respect to the spontaneous emission, in order to understand what are the conditions under which such mechanism can become significant in boosting the quasar RRL flux.

To this purpose, we take advantage of the formalism introduced by Shaver (1978). According to equation (1) of this work, the ratio $X$ between the stimulated $f_{\text{H}n\alpha}^{\text{stim}}$ and spontaneous emission $f_{\text{H}n\alpha}^{\text{spont}}$ can be written as follows:
\begin{equation}
X=\frac{f_{\text{H}n\alpha}^{\text{stim}}}{f_{\text{H}n\alpha}^{\text{spont}}}=\frac{c^2S_0}{2k_B\nu_{\text{obs}}^2\Omega T_e}\times {\cal F},
\label{xfac}
\end{equation}
where $k_B=1.38\times 10^{-16}$ erg/K is the Boltzmann constant, $\nu_{\text{obs}}=\nu/(1+z)$ is the observed frequency of the transition, $T_e$ is the electron temperature, $\Omega=(l/d)^2$ is the solid angle subtended by the H{\small II} region, $l$ is the total path length through the ionized gas and $d$ is the proper distance to the source, related to the luminosity distance $d_L$ through $d=d_L/(1+z)^2$, $S_0$ is the continuum flux density of the background radio source, and ${\cal F}$ is
\begin{equation}
{\cal F}=\Bigg[\frac{e^{-\tau_C}|e^{-\tau_L}-1|}{\Big(\frac{b_{m}\tau_L^*+\tau_C}{\tau_L+\tau_C}\Big)(1-e^{-(\tau_L+\tau_C)})-(1-e^{-\tau_C})}\Bigg]. 
\label{F}
\end{equation}
In eq. (\ref{F}), $b_m$ is the departure coefficient which relates the population of the atomic energy level $m=n+1$ (for $\alpha$-lines) to its value in local thermodynamic equilibrium (LTE), $\tau_C$ is the continuum optical depth and $\tau_L=b_n\beta_n\tau_L^*$ is the central line optical depth, where:
\begin{equation}
\beta_n\equiv 1-\frac{kT_e}{h\nu}\frac{\Delta b}{b_n}.
\end{equation}
The optical depths $\tau_C$ and $\tau_L^*$ can be written as
\begin{equation}
\tau_C=\frac{0.03014n_e^2lf}{\nu_{\text{obs}}^2T_e^{3/2}}[1.5\ln T_e-\ln (20.2\nu_{\text{obs}})]
\end{equation}
and\footnote{The expression for $\tau_L^*$ assumes that the profile width is due to pure Doppler broadening.}
\begin{equation}
\tau_L^*=\frac{575n_e^2lf}{\nu_{\text{obs}} \Delta V_{\text{obs}}T_e^{5/2}}x_{\text{HII}}e^{\frac{1.58\times10^5}{n^2T_e}}.
\end{equation}
Here $\Delta V_{\text{obs}}$ (km/s) is the observed FWHM of the line ($\Delta V_{\text{obs}}=\Delta V/(1+z)$), $n_e$ (cm$^{-3}$) is the electron density, $f(=V_{\text{gas}}/V_{\text{total}})$ is the volume filling factor, $x_{\text{HII}}$ is the ionized fraction of the gas and $n$ is the principal quantum number. Now, since $l=2R_{\text{HII}}$, where $R_{\text{HII}}^3=3\dot N_{\gamma}/(4\pi\alpha(T)n_e^2)$ is the Stromgren radius, eq. (\ref{xfac}) becomes:
\begin{equation}
X=\frac{3.10\times10^{-34}c^2\alpha(T)^{2/3}}{k_B\nu^2\dot N_{57}^{2/3}}\bigg(\frac{n_e^{4/3}S_0}{T_e}\bigg)\frac{d_L(z)^2}{(1+z)^2}\times \cal{F},
\label{xfac_final}
\end{equation}
where $\alpha$ (cm$^3$ s$^{-1}$) is the recombination rate, $\dot N_{57}$ (s$^{-1}$) is the rate of production of ionizing photons in units of $10^{57}$, and the units of $\nu$, $d_L$ and $S_0$ are GHz, Mpc and mJy respectively\footnote{Here $k_B$ is in units of erg/K and $c$ is in cm/s.}.\\

%In this work, we consider two components of the interstellar medium: H{\small II} regions ($n_e>10^3$ cm$^{-3}$; $T_e\approx 10^4$ K) and diffuse interclouds ($10^{-2} < n_e/\text{cm}^{-3} < 1$; $10^2<T_e/\text{K}<10^3$). These represent the extreme cases of interstellar clouds presented by Shaver (1975).\\
Following the formalism presented by Shaver (1975), the interstellar medium can be classified according to three main categories: HII regions ($10<n_e/\text{cm}^{-3}<10^4$; $2.5\times 10^3<T_e/\text{K}<10^4$), diffuse intercloud components ($n_e \sim 5\times 10^{-2}$ cm$^{-3}$; $T_e\sim 10^3$~K) and cold cloud components ($n_e \sim 5\times 10^{-2}$cm$^{-3}$; $T_e\sim 20$~K). With the aim of bracketing the physical properties of the ISM, in this work, we consider only two extreme cases: ($n_e>10^3$ cm$^{-3}$; $T_e\approx 10^4$ K) and ($10^{-2} < n_e/\text{cm}^{-3} < 1$; $10^2<T_e/\text{K}<10^3$), hereafter referred to as H{\small II} regions and diffuse interclouds, respectively.\\
In the case of H{\small II} regions, we assume LTE ($b_n=1$, $\beta_n=1$) and we compute the X factor for RRLs of quantum numbers detectable at $z=0$ and $z=6$ in correspondence of two reference SKA frequencies, i.e. the minimum and maximum observable with SKA-MID, as shown in Table \ref{table_SKA}. We can verify that in the LTE case, $\cal{F}$$=1$. 

\begin{table}%[h]
\caption{Quantum numbers of radio recombination lines detectable at $z=0$ and $z=6$ in correspondence of the two extreme SKA-MID frequencies.}
\centering
\begin{tabular}{lcc}
\hline
\hline
$\nu_{\text{obs}}$(GHz) & $n(z=0)$ & $n(z=6)$\\
\hline
\hspace{0.4cm}$0.35$ & $265$ & $139$\\
\hspace{0.4cm}$14$ & $77$ & $40$\\
\hline
\hline
\end{tabular}
\label{table_SKA}
\end{table}

In Figure \ref{contours_z0_z6} we show, at $z=0$ (left panel) and $z=6$ (right panel), contour plots of the logarithm of the X factor for these particular RRLs as a function of $T_e$ and $n_e$ in the intervals ($10^3<T_e/\text{K}<10^5$) and ($10^3<n_e/\text{cm}^{-3}<10^5$), for a radio continuum flux density\footnote{Radio-quiet quasars at $z\sim 6$ have typical radio fluxes of the order of $\lesssim$ tens of $\mu$Jy (Wang et al. 2008).} $S_0=10$ mJy and $S_0=0.01$ mJy respectively. Black solid contours are for RRLs detectable in correspondence of $\nu_{\text{min}}=0.35$ GHz, while red dashed contours correspond to $\nu_{\text{max}}=14$ GHz. We consider our ``fiducial'' case for quasars including secondary ionizations from $X$-ray photons, in which $\dot N_{57}=(8.76\times 10^{-11})\times10^{-0.4M_{\text{AB}}^{1500}}$, and we assume sources with $M_{\text{AB}}^{1500}=-26$.

\begin{figure}%[H]
\centering
\hspace{-0.7cm}
\includegraphics[width=0.51\textwidth]{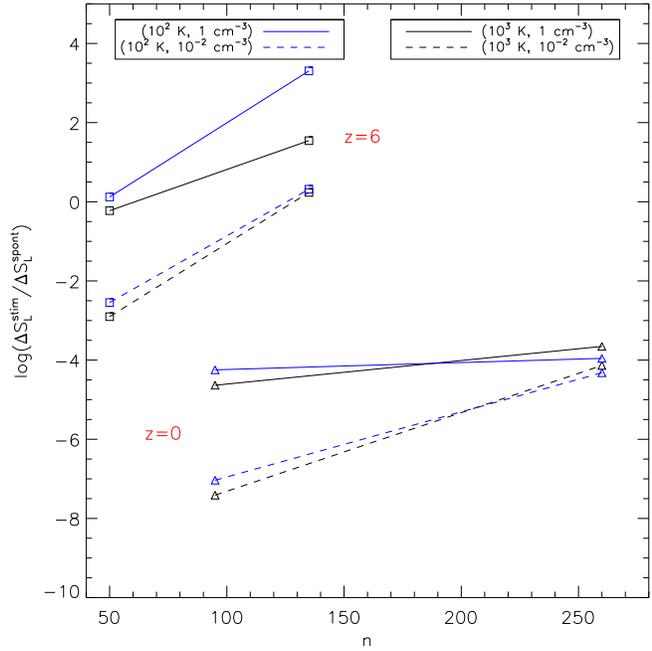}
\caption{Logarithm of the X factor for radio recombination lines of quantum numbers $n$ detectable at $z=0$ (triangles) and $z=6$ (squares) in correspondence of two SKA frequencies, $\nu=0.4$ GHz and $\nu=7.3$ GHz. The curves are plotted for two values of electron temperature: $T_e=10^2$ K (blue) and $T_e=10^3$ K (black), and two values of electron density: $n_e=10^{-2}$ cm$^{-3}$ (dashed) and $n_e=1$ cm$^{-3}$ (solid), both at $z=0$ and $z=6$. Our calculations assume $S_0=10^3$ mJy, $\Delta V=100$ km s$^{-1}$ and our ``fiducial'' case with secondary ionizations for a quasar magnitude $M_{\text{AB}}^{1500}=-26$.}
\label{frac_NLTE}
\end{figure}

We find that, at $z=0$ the logarithm of the X factor is in the interval $[-6,1]$, being greater than zero in the cases of very high-$n$ (i.e. $n=265$). This means that, except for these cases, the spontaneous emission dominates over the stimulated one for local sources. At $z=6$ instead, we find $-4 < \log X < 3$, being negative for very small quantum numbers. Thus, in all the situations but these, the stimulated radio flux due to the nonthermal continuum radiation can give a significant contribution to the total RRL flux density of the quasar at $z=6$. As a further comment, we notice that the X factor is linearly proportional to $S_0$, thus if $S_0$ reduces by a certain factor the stimulated contribution with respect to the spontaneous emission %$f_{\text{H}n\alpha}^{\text{stim}}$ 
also reduces by the same amount.
%scales with the radio flux density as $1/S_0$, implying that a variation of $S_0$ by a certain factor produces a variation of the stimulated contribution with respect to the spontaneous emission by the same amount.

In the case of diffuse interclouds, we cannot assume LTE. In this case, we use the $b_n$ and $\beta_n$ tabulated values as computed by Salem \& Brocklehurst (1979) for a nonthermal background source.\\
Furthermore, assuming the cloud is spherically symmetric ($f=1$) and the gas inside this region is fully ionized ($x_{\text{HII}}=1$), we rewrite $\tau_C$ and $\tau_L^*$ as follows:
\begin{equation}
\tau_C=\frac{0.12n_e^{4/3}\dot N_{57}^{1/3}}{\nu^2T_e^{3/2}\alpha(T)^{1/3}}\bigg[1.5\ln T_e-\ln\bigg(\frac{20.2\nu}{1+z}\bigg)\bigg](1+z)^2
\end{equation}
and
\begin{equation}
\tau_L^*=\frac{2.31\times10^{3}n_e^{4/3}\dot N_{57}^{1/3}}{\nu \Delta V T_e^{5/2}\alpha(T)^{1/3}}e^{\frac{1.58\times10^5}{n^2T_e}}(1+z)^2.
\end{equation}

In Figure \ref{frac_NLTE} we show the logarithm of the X factor for non-LTE cases for H$n\alpha$ lines as a function of the quantum numbers $n=95, 260$ at $z=0$ and $n=50, 135$ at $z=6$. We make the calculation in the case of quasars with $M_{\text{AB}}^{1500}=-26$ in our ``fiducial'' case including secondary ionizations, and assuming $S_0=10^3$ mJy and $\Delta V=100$ km s$^{-1}$. We plot the results for two values of electron temperature: $T_e=10^2$ K (blue curves) and $T_e=10^3$ K (black curves), and two values of electron density: $n_e=10^{-2}$ cm$^{-3}$ (dashed) and $n_e=1$ cm$^{-3}$ (solid) , both at $z=0$ and $z=6$.

We note that at $z=0$, log($X$) is between 4 and 5 orders of magnitude lower than the corresponding ratio (with the same $T_e$ and $n_e$) at $z=0$. Also, at a fixed density, log($X$) takes greater values for lower temperatures, while, at fixed $T_e$, it increases with increasing $n_e$; in all the cases, the ratio between the stimulated and spontaneous flux density increases with increasing $n$, more rapidly at $z=6$ with respect to $z=0$. We can conclude that, in the situation with low densities and low temperatures, the nonthermal stimulated emission dominates over the spontaneous emission only at $z=6$ and with $n_e\gtrsim 1$ cm$^{-3}$, for RRLs of quantum numbers $n\gtrsim 70$. Indeed, the contribution of the nonthermal radiation field becomes more important at the higher energy levels (or lower frequencies), with a more rapid increasing of log($X$) for $T_e=10^2$ K ($\log(X)\approx 3.5$ for $n=135$) with respect to the case of $T_e=10^3$ K ($\log(X)\approx 1.5$ for $n=135$).
%\\{\bf (Last paragraph removed)}

%In conclusion, we can say that, in our ``fiducial'' case with the inclusion of secondary ionizations from X-rays, the stimulated emission due to a nonthermal continuum radio source can give a significant contribution to the RRL emission from quasars with $M_{\text{AB}}^{1500}=-26$ in the following considered situations:
%\begin{enumerate}
%\item at $z=0$, in the case of high electron temperatures ($10^3$ K $<T_e<10^5$ K) and high electron densities ($10^3$ cm$^{-3}<n_e<10^5$ cm$^{-3}$), for $n=265$, $S_0=10$ mJy ($-2<\log X<0$).
%\item At $z=6$, in the case of high electron temperatures ($10^3$ K $<T_e<10^5$ K) and high electron densities ($10^3$ cm$^{-3}<n_e<10^5$ cm$^{-3}$), for $n=139, S_0=0.01$ mJy ($-4<\log(X)<0$); in the case of low electron temperatures ($10^2$ K $<T_e<10^3$ K) and electron densities $n_e\gtrsim 1$ cm$^{-3}$, for $n\gtrsim 70$, $S_0=10^3$ mJy ($-3.5<\log(X)<0$).
%\end{enumerate}

\begin{figure}%[H]%[ht!]
\centering   
\hspace{-1.03cm}   
 \includegraphics[width=0.53\textwidth,height=0.33\textheight]{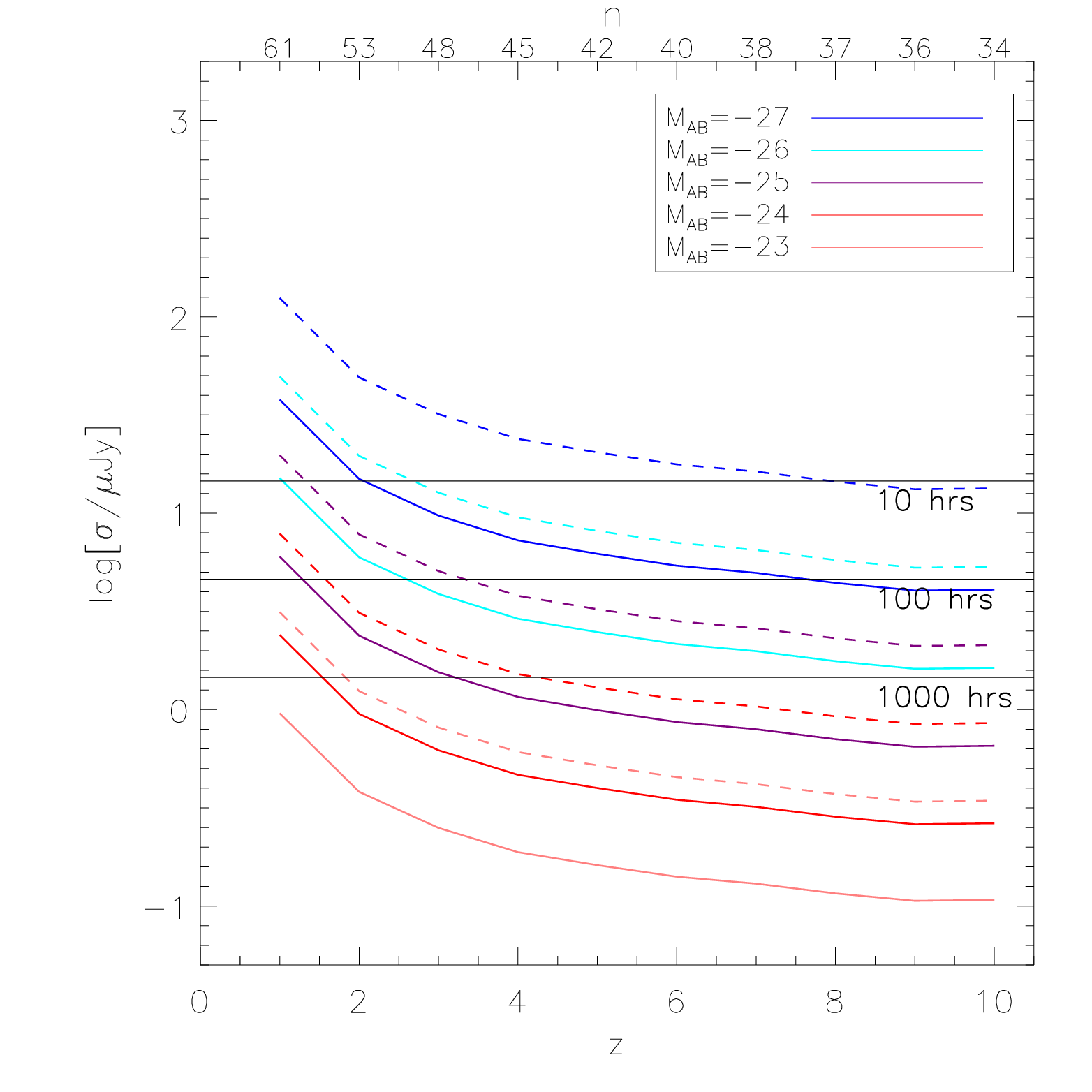}
\caption{Logarithm of the 5$\sigma$ sensitivity (in $\mu\text{Jy}$) as a function of redshift ($1 \leq z \leq 10$) for different values of the quasar absolute magnitude ($-27 \leq M_{\text{AB}}^{1500} \leq -23$). The calculations are done for the maximum frequency observable with SKA-MID, $\nu_{\text{max}}=14$ GHz. The solid and dashed curves correspond to our ``fiducial'' and ``extreme'' model respectively, with the inclusion of the stimulated contribution from a nonthermal radio source, using the mean value of $\log(X)$ obtained for $S_0=10$ $\mu$Jy at this frequency. The horizontal black lines represent the 5$\sigma$ ``threshold'' sensitivities reached by SKA-MID at the corresponding frequency, with an integration time of 10, 100 and 1000 hours. In addition, the upper $x$-axis indicates the values of the quantum numbers $n$ of the lines detectable at each $z$, corresponding to $\nu_{\text{max}}$. The H$n\alpha$ line intensity decreases with decreasing frequency.}
   \label{sigma_z}
\end{figure}

\section{Detecting (obscured) quasars}
\hspace{12pt}
Given the model outlined in Sec. 2 and taking into account the stimulated emission due to a nonthermal background source, we estimate the expected H$n\alpha$ line flux density arising from quasars of different AB magnitudes ($-27 \leq M_{\text{AB}}^{1500} \leq -23$) and redshifts ($1 \leq z \leq 10$), both in the ``fiducial'' and ``extreme'' case\footnote{We assume $\delta v=100$ km $\text{s}^{-1}$ in these calculations.}, and we compute the sensitivity ($\sigma=f_{\text{H}n\alpha}/5$) required to detect H$n\alpha$ lines with 5$\sigma$ significance at all considered redshifts and AB magnitudes. The total (spontaneous + stimulated) RRL flux density is
\begin{equation}
f_{\text{H}n\alpha}^{\text{tot}}=f_{\text{H}n\alpha}(1+X),
\label{fHnalpha_TOT}
\end{equation}
where $f_{\text{H}n\alpha}$ includes only the spontaneous emission from quasars, and $X$ is the ratio between stimulated and spontaneous flux density from eq. (\ref{xfac_final}). In our calculations, we use the mean value of $\log X$ obtained for RRLs observable in correspondence of the maximum frequency observable with SKA-MID, $\nu_{\text{max}}=14$ GHz, for a radio continnum flux\footnote{The assumption $S_0=10$ $\mu$Jy underestimates the contribution of the stimulated emission in low-redshift quasars, whose observed radio fluxes are typically higher.} $S_0=10$ $\mu$Jy.

The results are shown in Fig. \ref{sigma_z}, for the concerned ranges of magnitude and redshift, in the ``fiducial'' (solid curves) and ``extreme'' (dashed curves) case.\\
In this figure we also plot, through horizontal black lines, the 5$\sigma$ SKA sensitivity (5$\sigma_{\text{SKA}}$) that is expected to be reached in $t=10$, $100$ and $1000$ hours of SKA-MID observing time at the corresponding frequency. We adopt the usual formula:
\begin{equation}
\sigma_{\text{SKA}}=\sigma_{0}\sqrt{\frac{t_{0}}{t}\frac{\Delta v_{0}}{\Delta v}},
\end{equation}
where $\sigma_{0}=63\hspace{0.1cm}\mu\text{Jy}$ is the sensitivity obtained over a channel $\Delta \nu_{0}=100$ kHz in $t_{0}=1$ hr of SKA-MID integration time (SKA-­‐SCI-­‐LVL-­‐001, 2015), $\Delta v_{0}=c\Delta \nu_{0}/\nu _{0}$, $\Delta v=100$ km $\text{s}^{-1}$ and $\nu_{0}=\nu_{\text{max}}$. 

Considering that: a) $f_{\text{H}n\alpha}$ increases with increasing frequency, and b) the stimulated contribution increases (i.e. $X$ increases) with decreasing frequency, at $\nu=\nu_{\text{max}}$ we are maximizing the expected H$n\alpha$ flux but we are minimizing the stimulated emission.

From the figure, we can see that at $\nu_{\text{max}}=14$ GHz, and according to our ``fiducial'' model, SKA-MID could detect sources with $M_{\text{AB}} < -26$ ($M_{\text{AB}} < -25$) at all redshifts (at $z\lesssim 3$) in 1000 hrs of integration time. In the ``extreme'' case, quasars up to $M_{\text{AB}}=-25$ ($M_{\text{AB}}=-24$) at all redshifts (at $z\lesssim 4$) would be detected in the same (large) amount of observing time. In 100 hrs of observing time, SKA-MID would detect sources with $M_{\text{AB}}<-27$ ($M_{\text{AB}}<-26$) at $z\lesssim 8$ ($z\lesssim 2.5$) in our ``fiducial'' case, and sources with $M_{\text{AB}}<-26$ ($M_{\text{AB}}<-25$) at all redshifts (at $z\lesssim 3$) in the ``extreme'' case.\\
As stated in Sec. 3, our predicted flux densities are $\sim 2$ orders of magnitude below the observed fluxes of local AGNs (Figure \ref{flux_densities}). 
%It still typically require hundreds of hours on a source, which is unlikely to happen except for some very deep small area surveys and maybe some special sources. However, I think this calculation is based on your calculations - right? 
This means that, if some relevant contribution to the RRL emission is still missing, then the integration time would go down by a factor $>10$, making RRLs from quasars much more detectable.

Furthermore, Fig. \ref{sigma_z} shows that moving from $z\sim 1$ to $z\sim 10$, the sensitivity required for detecting quasars through RRLs changes only by a factor of 10. To understand this trend in $z$, and since $f_{\text{H}n\alpha} \propto (1+z)d_L^{-2} n^{-2.72}$ we also express $n$ and $d_L$ as a function of redshift. For a flat universe ($1-\Omega_m-\Omega_\Lambda=0$), $d_L$ is 
\begin{equation}
d_L(z)=\frac{c}{H_0}(1+z)\int_0^z\frac{dz'}{[\Omega_m(1+z')^3+\Omega_\Lambda]^{1/2}},
\label{dL}
\end{equation} 
where $H_0$ is the Hubble parameter, and $\Omega_m$ and $\Omega_\Lambda$ are the total matter density and the dark energy density in the units of critical density, respectively. At high redshift, by neglecting $\Omega_{\Lambda}$, we obtain that $d_L \propto (1+z)$. The quantum number $n$ can be written in terms of $z$ by using eq. (\ref{nu_obs}) and making some approximations for high $n$: the resulting relation is
\begin{equation}
n \approx \left(\frac{2c\hspace{0.05cm}\text{R}_{\text{H}}}{\nu_{\text{obs}}(1+z)}\right)^{1/3}.
\label{n}
\end{equation}
We finally get $f_{\text{H}n\alpha} \propto (1+z)^{-0.1}$, an expression that reproduces the slow decrease of $\sigma$ with increasing redshift.

%{\bf Moreover, we add some discussion about what would be the ``optimal'' frequency/range which maximizes the expected RRL flux from quasars. As said above, the two factors which play a role in determining the total flux $f_{\text{H}n\alpha}^{\text{tot}}$ are the spontaneous term and the boosting factor $X$; their frequency dependence is respectively: $f_{\text{H}n\alpha}^{\text{spont}}\propto \nu^{0.91}$ (from eq. (\ref{fHnalpha}) with the approximation given by eq. (\ref{n})), and $X\propto \nu^{-2}$ (eq. (\ref{xfac_final})). Thus, given that the boost is more strongly dependent on $\nu$ with respect to $f_{\text{H}n\alpha}^{\text{spont}}$, a large increase in frequency which on one hand produces a quite large increase in $f_{\text{H}n\alpha}^{\text{spont}}$, on the other hand causes a decrease in $X$ which is dominant with respect to the increasing term. This implies that a too large increase in $\nu$ (e.g. up to 100 GHz) is not optimal for maximizing the total flux density, i.e. a frequency $\nu\sim 14$ GHz is nearly ideal.}
%we quantify the frequency dependence of both the spontaneous H$n\alpha$ flux density and the factor $X$

The optimal strategy to detect RRLs from quasars depends on the nature of the source itself. As discussed above, the two factors which play a role in determining the total flux $f_{\text{H}n\alpha}^{\text{tot}}$ are the spontaneous term ($f_{\text{H}n\alpha}^{\text{spont}}\propto \nu_{\text{obs}}^{0.9}$) and the stimulated one ($f_{\text{H}n\alpha}^{\text{stim}}\propto \nu_{\text {obs}}^{-1}$). This means that, in the case of a radio-quiet source ($f_{\text{H}n\alpha}^{\text{stim}}<<f_{\text{H}n\alpha}^{\text{spont}}$), high frequencies (e.g. $\nu_{\text{obs}}\sim 14$~GHz in the case of SKA-MID) optimize the chance of RRLs detection. Viceversa, in the case of a radio-loud source it is better to look for RRLs by means of low-frequency receivers (e.g. $\nu_{\text{obs}}\sim 1$~GHz in the case of SKA-MID).
\section{Summary and discussion}
\hspace{12pt}We have examined the feasibility of detecting obscured quasars at radio frequencies, through their hydrogen RRL emission. To this purpose, we have developed a model for estimating the expected flux densities of H$n\alpha$ lines from quasars, including secondary ionizations from X-ray photons.
%\begin{itemize}

By comparing our results with observations of RRLs in local quasars, we have found that our predictions are in good agreement with the observed RRLs strengths. Our predicted fluxes are, however, still below the observed line fluxes for local AGNs by a factor of a few orders of magnitude, suggesting that some additional mechanism is at work to amplify line fluxes. We have shown that, under some specific conditions, the contribution of stimulated emission due to a nonthermal radio source may be relevant in boosting the flux densities from quasars, thus making possible the detection of RRLs out to very large distances. Additionally, we have explored the potential of the SKA in detecting quasars at very high redshift ($z>7$). We have found that, taking into account the stimulated contribution, at a frequency $\nu=14$ GHz, the SKA-MID telescope would be able to detect sources with $M_{\text{AB}}\lesssim-27$ ($M_{\text{AB}}\lesssim-26$) at $z\lesssim 8$ ($z\lesssim 3$) in less than 100 hrs of integration time.
%\end{itemize}
%\\{\bf (bullets removed)}

We point out that, if we study the possible detectability of radio recombination lines based on the observed flux densities in local AGNs, which are brighter than our predictions by $\sim 2$ orders of magnitudes, then the required observing time would reduce by a factor $>10$, thus making RRLs from quasars more easily detectable.

Given our predictions, the proposed survey could be carried out as a piggyback of proposed deep HI surveys with SKA-MID (Blyth et al. 2015).

It must be noticed that we are maximizing the expected H$n\alpha$ flux through our assumptions $f_{\text{esc}}^{912}=0$ in eq. (\ref{main_formula}) and $x_e=0$ in eq. (\ref{sec_ion}). By assuming that ionizing photons can not escape from the galaxy, being trapped into the dense ISM of the quasar host galaxy, we are maximizing the number of ionizations (and consequent recombinations) in HII regions. Moreover, by considering the case of an almost complete neutral gas, we are maximizing the amount of free-electrons energy that can be used to produce secondary ionizations. 

We remark here that the possibility of detecting RRLs even at high redshift is due to an important property of these lines, concerning with the fact that their principal quantum number decreases with increasing redshift (see Section 1 and Figure \ref{nuobs_z}). 

Furthermore, we have obtained a relation between the quantum number of a RRL detectable at a given $z$ in correspondence of some observed frequency: $n \approx (2c\hspace{0.05cm}\text{R}_{\text{H}}/[\nu_{\text{obs}}(1+z)])^{1/3}$, valid for high $n$. This explains the slow decrease of the H$n\alpha$ line flux density with increasing redshift ($f_{\text{H}n\alpha} \propto (1+z)^{-0.1}$).

Also, it is worthwhile to mention a recent result from Ba$\tilde{\text{n}}$ados et al. (2015), showing that the fraction of radio-loud quasars (i.e. RLF $= 6\%-19\%$) does not vary with redshift, in contrast to what previously found by some studies at lower redshifts (e.g. Jiang et al. 2007; Kratzer \& Richards 2014).

Our results for the line flux densities must be still considered as preliminary. More detailed radiative transfer computations of UV/Xray photons performed on realistic density field are necessary to draw firmer conclusions. We aim at presenting these improved results in a future work, where we plan to use the state-of-the-art RT code \texttt{CRASH} in its latest version (\texttt{CRASH4}, Graziani et al. in prep). The ratio between the intensity of the $\text{He}^{++}$ and $\text{H}^{+}$ lines is expected to be a good diagnostic of the hardness of the ionizing spectrum, thus providing a clear discriminant between AGN and starburst activity. The advantage of this technique is that both lines can be detected contemporarily with a bandwidth of 10 GHz (Scoville $\&$ Murchikova 2013). We will investigate how this ratio varies for different slopes of the quasar spectrum and for different values of the ionization fraction of the gas. Since this technique can be applied also in the millimetre domain, the proposed scientific case can be, on longer time scales, a driver for the synergy between the SKA and ALMA.

Finally, we also plan to study the redshift evolution of the UV luminosity function, by fitting the most updated observational data at $0<z<7$, and theoretically extending it beyond $z=7$. This will allow to make detailed predictions of the expected QSO number counts for a single observation with SKA-MID (FoV = 0.49 $\text{deg}^2$) in 1000 hrs of integration time, taking into account the full frequency coverage of this instrument (0.35-14 GHz).

\section*{Aknowledgements}
\hspace{12pt}This research has made use of the NASA/IPAC Extragalactic Database (NED) which is operated by the Jet Propulsion Laboratory, California Institute of Technology, under contract with the National Aeronautics and Space Administration.

This work is based on research supported by the National Research Foundation under grant 92725. Any opinion, finding and conclusion or recommendation expressed in this material is that of the author(s) and the NRF does not accept any liability in this regard.

%{\bf (References are now up-to-date)}

\end{document}